\documentclass[aps,showpacs,preprint]{revtex4}
\usepackage{amstext}
\usepackage{amssymb}
\usepackage{graphicx}

\begin{document}

\title{A new derivation of symmetry energy from nuclei beyond the $\beta$%
-stability line}
\author{V.M.~Kolomietz and A.I.~Sanzhur}
\affiliation{Institute for Nuclear Research, 03680 Kiev, Ukraine}

\begin{abstract}
We suggest the procedure of direct derivation of the symmetry energy from
the shift of \ neutron-proton chemical potentials $\Delta \lambda =\lambda
_{n}-\lambda _{p}$ for nuclei beyond the beta-stability line. We observe the
presence of anomalous strong (about $15\%$) shell oscillations at the
symmetry energy coefficient $b_{\mathrm{sym}}$. Our results do not confirm
the existence of exceptionally large values of the symmetry energy
coefficient at mass number $A\approx 100\ $which was earlier\ reported in
Ref. \cite{jado03}. Using the fitting procedure, we have evaluated the
volume, $b_{\mathrm{sym,vol}}$, and surface, $b_{\mathrm{sym,surf}}$,
contributions to the symmetry energy. We have estimated the experimental
value of surface-to-volume ratio as $r_{S/V}=|b_{\mathrm{sym,surf}}|/b_{%
\mathrm{sym,vol}}\approx 1.7$\ for the fitting interval $A\geq 50$.
\end{abstract}

\maketitle

\textbf{1.} The nuclear $\beta $-stability line is derived by the balance of
both the isotopic symmetry, $E_{\mathrm{sym}}$, and the Coulomb, $E_{C}$,
energies. However the extraction of $E_{\mathrm{sym}}$ and $E_{C}$ from the
nuclear binding energy is not a simple problem because of its complicated
dependency on the mass number $A$ in finite nuclei \cite{auwa03}. The
standard procedure of extraction of the symmetry energy from a fit of mass
formula to the experimental binding energies \cite{jado03} is not free from
ambiguities and does not allow one to separate the symmetry energy into the
volume, surface and curvature contributions directly.

Moreover the symmetry energy $E_{\mathrm{sym}}$ is usually derived on the $%
\beta $-stability line and some special efforts have to be applied to extend
it beyond the ground state of nuclei \cite{onda04}. On the other hand, new
information about nuclear masses in a wide region of the stability valley
can be used for straightforward derivation of the $A$-dependence of \
energies $E_{\mathrm{sym}}$ and $E_{C}$.

In the present work, we suggest a non-standard procedure of extraction of
the symmetry and Coulomb energies from the experimental data using the
dependence of the\ isospin shift of \ neutron-proton chemical potentials $%
\Delta \lambda (X)=\lambda _{n}-\lambda _{p}$\ on the asymmetry parameter $%
X=(N-Z)/(N+Z)$\ for nuclei beyond the beta-stability line. This procedure
allows one to represent the results for the $A$-dependence of energies $E_{%
\mathrm{sym}}$\ and $E_{C}$\ in a transparent way, which can be easily used
for the extraction of the smooth volume and surface contributions as well as
the shell structure.

\textbf{2.} Considering the asymmetric nuclei with a small asymmetry
parameter $X=(N-Z)/A\ll 1$ and assuming the leptodermous property, the total
energy per nucleon $E/A$ can be represented in the following form of $A,X$%
-expansion 
\begin{equation}
E/A\equiv e_{A}=e_{0}(A)\ +b_{\mathrm{sym}}(A)\ X^{2}+E_{C}(X)/A,
\label{E/A}
\end{equation}%
where $e_{0}(A)$ includes both the bulk and the surface energies, $b_{%
\mathrm{sym}}(A)$ is the symmetry energy, $E_{C}(X)$ is the total Coulomb
energy 
\begin{equation}
E_{C}(X)=\frac{3}{20}\frac{Ae^{2}}{R_{C}}(1-X)^{2}  \label{eC}
\end{equation}%
and $R_{C}$ is the Coulomb radius of the nucleus.

The beta-stability line $X=X^{\ast }(A)$ can be directly derived from Eq. (%
\ref{E/A}) using the condition%
\begin{equation}
\left. \frac{\partial E/A}{\partial X}\right\vert _{A}=0\quad
\Longrightarrow \quad X^{\ast }(A)=\frac{\ e_{C}^{\ast }(A)}{b_{\mathrm{sym}%
}^{\ast }(A)+e_{C}^{\ast }(A)\ },  \label{cond}
\end{equation}%
where%
\[
e_{C}(A)=0.15Ae^{2}/R_{C} 
\]%
Along the beta-stability line, the binding energy per particle is then given
by%
\begin{equation}
E^{\ast }/A=e_{0}^{\ast }(A)\ +b_{\mathrm{sym}}^{\ast }(A)\ X^{\ast
2}+E_{C}(X^{\ast })/A,  \label{stab}
\end{equation}%
where the upper index \textquotedblright $\ast $\textquotedblright\
indicates that the corresponding quantity is determined by the variational
conditions (\ref{cond}) taken for fixed $A$ and $X=X^{\ast }$ on the
beta-stability line. For any given value of $A$, the binding energy can be
extended beyond the beta-stability line as 
\begin{equation}
E/A=E^{\ast }/A+b_{\mathrm{sym}}^{\ast }(A)(X-X^{\ast })^{2}+\Delta
E_{C}(X)/A,  \label{eq1}
\end{equation}%
where $\Delta E_{C}(X)=E_{C}(X)-$ $E_{C}(X^{\ast })$. The symmetry energy $%
b_{\mathrm{sym}}^{\ast }(A)$ contains the $A$-independent bulk term, $b_{%
\mathrm{sym,vol}}^{\ast }$, and the $A$-dependent surface contribution, $b_{%
\mathrm{sym,surf}}^{\ast }A^{-1/3}$,

\begin{equation}
b_{\mathrm{sym}}^{\ast }(A)=\ \ b_{\mathrm{sym,vol}}^{\ast }+b_{\mathrm{%
sym,surf}}^{\ast }\ A^{-1/3}.  \label{bi}
\end{equation}%
In general, the surface symmetry energy $b_{\mathrm{sym,surf}}^{\ast
}A^{-1/3}$ includes also the high order curvature correction $\sim $ $%
A^{-2/3}$ \cite{kosa08}.

Using Eq. (\ref{eq1}), one can establish an important relation for the
chemical potential $\lambda _{q}$ ($q=n$ for a neutron and $q=p$ for a
proton) beyond the beta-stability line. Namely, for the fixed $A$, we obtain
the following result\ from Eqs. (\ref{E/A}) and (\ref{stab}) 
\begin{equation}
\Delta \lambda (X)=\lambda _{n}-\lambda _{p}=\left. \frac{\partial E}{%
\partial N}\right\vert _{Z}-\left. \frac{\partial E}{\partial Z}\right\vert
_{N}=2\left. \frac{\partial (E/A)}{\partial X}\right\vert _{A}=4\left[ b_{%
\mathrm{sym}}^{\ast }(A)+e_{C}^{\ast }(A)\right] (X-X^{\ast }),
\label{lambdaX}
\end{equation}%
\ where 
\begin{equation}
\lambda _{n}=\left( \frac{\partial E}{\partial N}\right) _{Z}\ ,\ \ \
\lambda _{p}=\left( \frac{\partial E}{\partial Z}\right) _{N}\ .
\label{dlamb}
\end{equation}

On the beta-stability line, it follows Eq. (\ref{lambdaX}) that $\Delta
\lambda (X)_{X=X^{\ast }}=0$, as it should be from the definition of the
beta-stability line. We point out that for finite nuclei, the condition $%
\Delta \lambda =0$ on the beta-stability line is not necessary fulfilled
explicitly because of the discrete spectrum of the single particle levels
for both the neutrons and the protons near Fermi surface.

\textbf{3.} The quantity $\partial (E/A)/\partial X$ in Eq. (\ref{lambdaX})
can be evaluated within the accuracy of $\sim 1/A^{2}$ using the finite
differences which are based on the experimental values of the binding energy
per nucleon $\mathcal{B}(N,Z)=-E(N,Z)/A$. Namely, 
\begin{equation}
\left. \frac{\partial (E/A)}{\partial X}\right\vert _{A}=\frac{A}{4}\,\left[ 
\mathcal{B}(N-1,Z+1)-\mathcal{B}(N+1,Z-1)\right] .  \label{diff}
\end{equation}%
Since the difference (\ref{diff}) is taken for $\Delta Z=-\Delta N=2$, the
pairing effects do not affect the resulting accuracy.

From the binding energy tables \cite{auwa03} we have obtained sets of values
for $\Delta \lambda (X,A=\mathrm{const})$ covering mass numbers from $8$ to $%
238$. Each set contains $3$ to $11$ points $(X,\Delta \lambda /4)$. In 
\textrm{Fig.~1}, the typical sets of values are plotted for $A=100$, $120$
and $160$. As seen from \textrm{Fig.~1}, the positions of symbols determined
by (\ref{diff}) can be reproduced quite well by the linear dependence on $X$%
. This allows one to extract values of quantities $b_{\mathrm{sym}}^{\ast
}(A),~e_{C}^{\ast }(A)$ and $X^{\ast }$ for a given mass number $A$, see
Eq.~(\ref{lambdaX}). The linear functions $\Delta \lambda (X)$ from the Eq. (%
\ref{lambdaX}) obtained by the best fit to the experimental data are shown
by solid lines in \textrm{Fig.~1}.

\vspace*{3ex}
%
%  FIGURE 1
%
\includegraphics[width=0.8\textwidth]{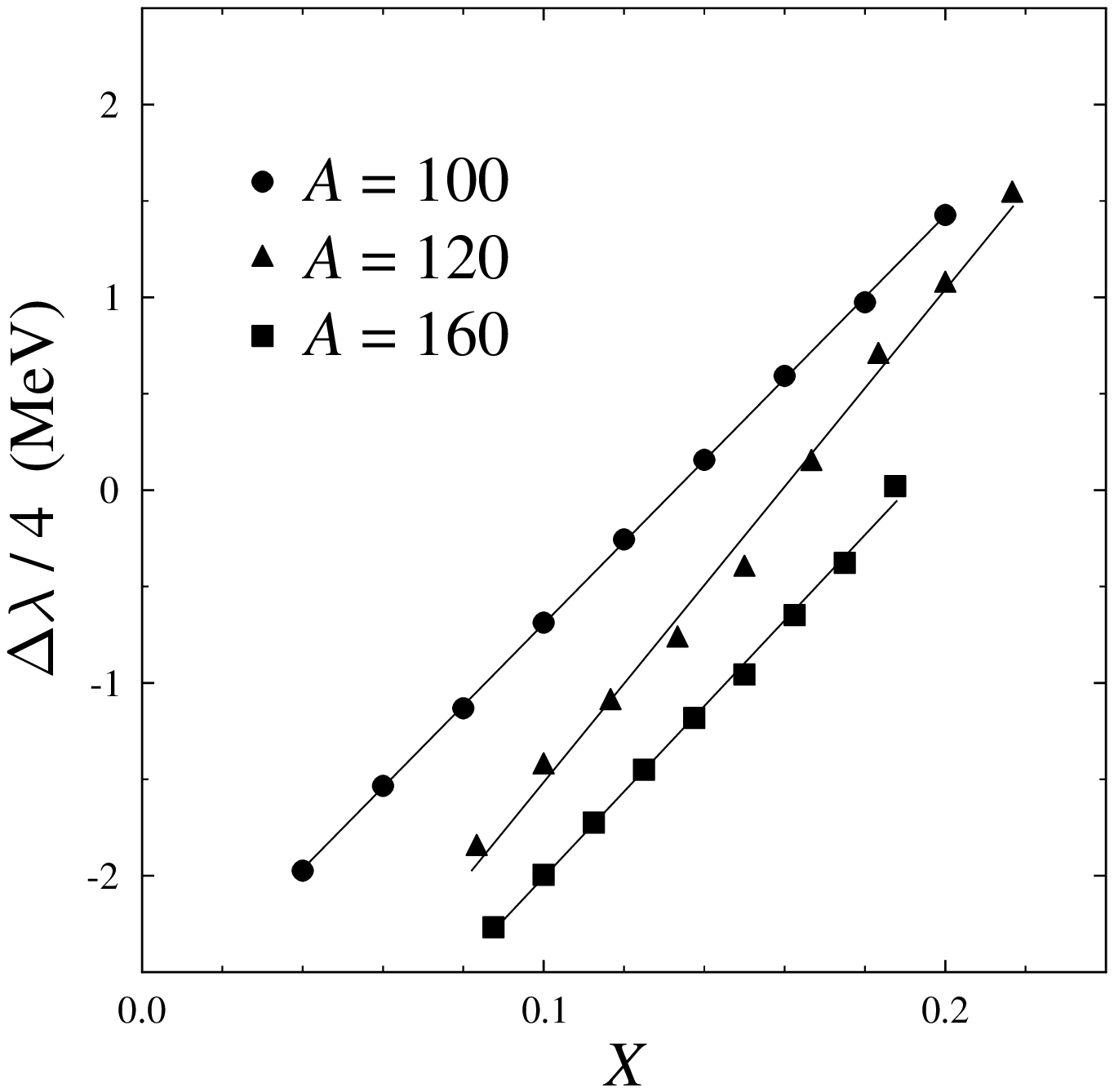}\newline
{\small Fig. 1. The difference $\Delta \lambda $ between neutron and proton
chemical potentials as a function of the asymmetry parameter $X$ for nuclei
with mass numbers $A=100,\ 120$ and $160$. The experimental values are
presented by the symbols. Solid lines show the result for $\Delta \lambda /4$
using Eq. ({\ref{lambdaX}}) with values $b_{\mathrm{sym}}^{\ast }$ and $%
e_{C}^{\ast }(A)$ obtained from the least square fit.}

In agreement with the Eq. (\ref{lambdaX}), the slopes of straight lines in 
\textrm{Fig. 1} allow us to derive the quantity $b_{\mathrm{sym}}^{\ast
}(A)+e_{C}^{\ast }(A)$. From the beta-stability condition $\Delta \lambda
(X)=0$ one can also derive the asymmetry parameter $X^{\ast }(A)$. Finally,
using the Eq. (\ref{cond}), we obtain the symmetry energy coefficient $b_{%
\mathrm{sym}}^{\ast }(A)$ and Coulomb energy parameter $e_{C}^{\ast }(A)$.
The corresponding results are shown in \textrm{Figs. 2, 3} and \textrm{4}.
Note that the error bars in \textrm{Figs.~2, 3} and \textrm{4} represent the
standard error interval obtained for\textbf{\ }$b$\textbf{$_{\mathrm{sym}%
}^{\ast }$}, $e_{C}^{\ast }(A)$ and $X^{\ast }$ from the least square fit
for the corresponding set $\Delta \lambda (X,A=\mathrm{const})$. These
errors reflect the systematic deviation of $\Delta \lambda (X)$ from the
linear dependence. Experimental errors in the nuclear masses are about on
order of magnitude less than that shown by the error bars. The error bars,
however, are small enough to see the non-monotonic behavior (the shell
structure) of the plots versus mass number.

\vspace*{2ex}
%
%  FIGURE 2
%
\includegraphics[width=0.85\textwidth]{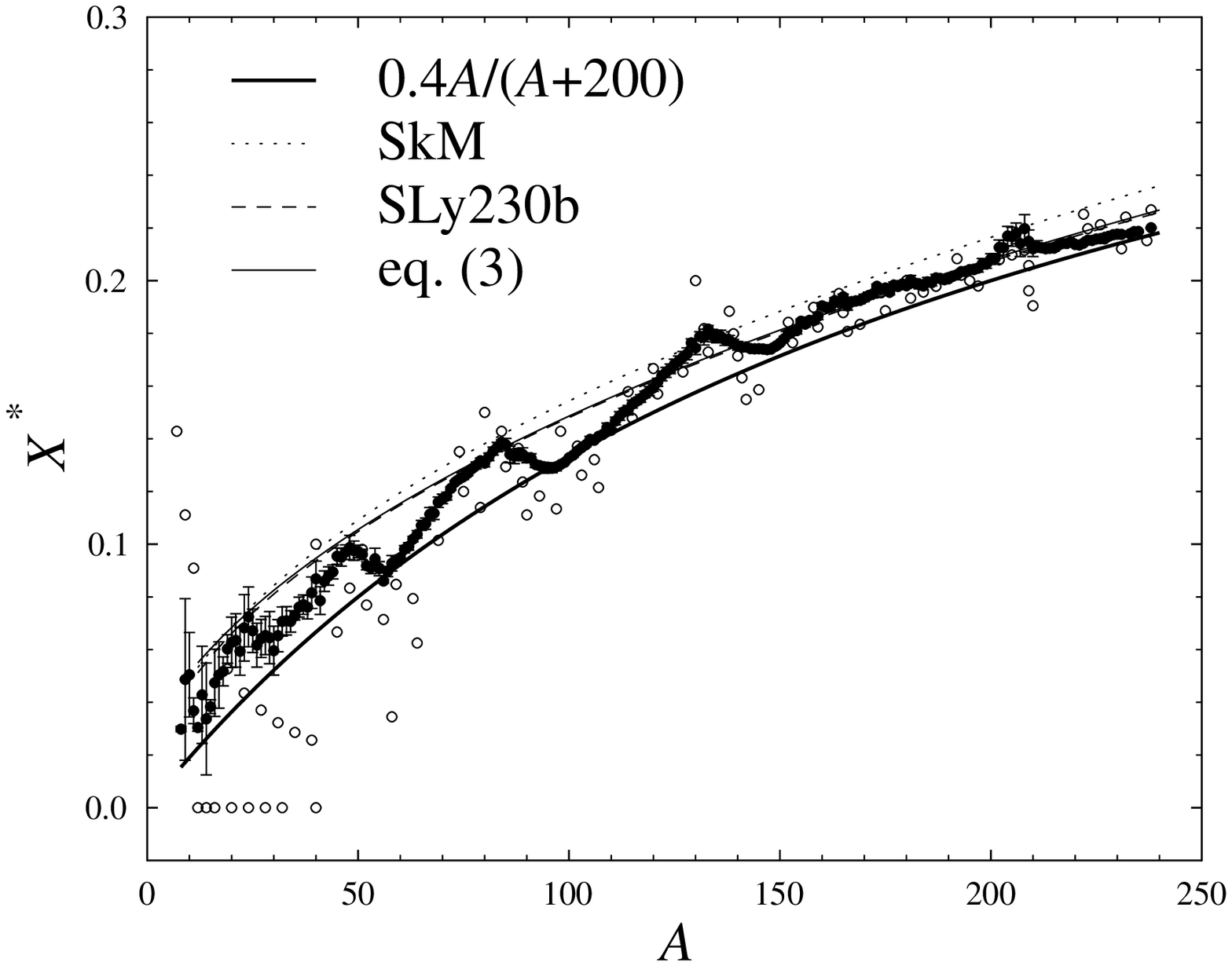}\newline
{\small Fig. 2.\ Asymmetry parameter $X^{\ast }(A)$ versus the mass number $A$.
Filled circles -- experimental data, dashed lines -- calculations using
Skyrme forces SkM and SLy230b, see Ref. \cite{kosa08}. Solid lines~--
$X^{\ast }(A)=0.4A/(A+200)$ \cite{gren53} (thick), and $X^{\ast}(A)=
e_{C}^{\ast }(A)/\left[ b_{\mathrm{sym}}^{\ast }(A)+e_{C}^{\ast}(A)\right]$
(see (\ref{cond})) with $e_{C}^{\ast }(A)=0.17A^{2/3}$,
$b_{\mathrm{sym}}^{\ast}(A)=26.5-25.6A^{-1/3}$ (thin). Open circles --
beta-stability line obtained from the periodic system of elements.}

In \textrm{Fig. 2}, we have plotted the value of $X^{\ast }(A)$ (solid dots)
which was obtained from Eq. (\ref{lambdaX}) and \textrm{Fig. 1}. The dashed
and dotted lines in \textrm{Fig. 2} were derived from the extended
Thomas-Fermi approximation for SkM and SLy230b Skyrme forces, respectively,
see Ref. \cite{kosa08}. The solid (thick) line in \textrm{Fig. 2} was
obtained using the empirical formula \cite{gren53} 
\begin{equation}
X^{\ast }(A)=\frac{0.4\ A}{A+200}.  \label{xstar}
\end{equation}%
The "experimental" curve $X^{\ast }(A)$ in \textrm{Fig. 2 } shows the
non-monotonic\ (sawtooth) shape as a function of the mass number $A$. This
behavior is the consequence of shell structure of single particle levels
near Fermi surface for both the neutrons and the protons. Because of this
shell structure, the Fermi levels for protons and neutrons can coincide by
chance only. For light nuclei, this significantly affects the beta-stability
line, deflecting it from the empirical one for stable nuclei (open circles
in \textrm{Fig. 2}). In agreement with \ Eq. (\ref{cond}), the smooth
behavior of $X^{\ast }(A)$ is achieved by a fit of the symmetry and the
Coulomb energy coefficients. Thin solid line in \textrm{Fig. 2} is obtained
from Eq. (\ref{cond}) with $e_{C}^{\ast }(A)=0.17A^{2/3}$ and $b_{\mathrm{sym%
}}^{\ast }(A)=26.5-25.6A^{-1/3}$.

The $A$-dependence of the Coulomb energy coefficient $e_{C}^{\ast }(A)$ is
shown in \textrm{Fig.~3}. Assuming $R_{C}\propto A^{1/3}$, one obtains that $%
e_{C}^{\ast }(A)\propto A^{2/3}$. Using this $A$-dependence of $e_{C}^{\ast
}(A)$, we have the best fit to the experimental values by using $e_{C}^{\ast
}(A)=0.17A^{2/3}$. The corresponding result of the fit is shown in \textrm{%
Fig.~3} as the solid line. The deviation of the Coulomb energy coefficient
from the smooth $A$-dependence is mainly due to the shell oscillations of
the Coulomb radius $R_{C}$.

The dependency of the symmetry energy coefficient $b_{\mathrm{sym}}^{\ast }$
on the mass number $A$ obtained from the experimental nuclear masses using
Eqs. (\ref{lambdaX}) and (\ref{diff}) is shown in \textrm{Fig.~4} as solid
circles. This dependence shows the strong shell oscillations with the
amplitude of about $15\%$. For the purpose of comparison, one could recall
that shell effects contribute about $1\%$ to the nuclear mass. In this
paper, we have performed the fit of experimental data for $b_{\mathrm{sym}%
}^{\ast }$ to the leptodermous-like functional form of Eq. (\ref{bi}). To
extract the expansion coefficients $b_{\mathrm{sym,vol}}^{\ast }$ and $b_{%
\mathrm{sym,surf}}^{\ast }$ from the fit we have used the data with $A\geq
12 $ for which the justified leptodermous expansion. Assuming $b\mathbf{_{%
\mathrm{sym}}^{\ast }}(A)$ given by Eq.~(\ref{bi}) as the basic dependency
of the symmetry energy coefficient on the mass number, we have obtained $b_{%
\mathrm{sym,vol}}^{\ast }=26.5$~\textrm{MeV} and $b_{\mathrm{sym,surf}%
}^{\ast }=-25.6$~\textrm{MeV }with the surface-to-volume ratio\textbf{\ }$%
r_{S/V}=|b_{\mathrm{sym,surf}}^{\ast }|/b_{\mathrm{sym,vol}}^{\ast }\approx
1 $. The corresponding function $b_{\mathrm{sym}}(A)$ is plotted as the
solid line in \textrm{Fig.~4}. The calculated values of the coefficients $b_{%
\mathrm{sym,vol}}^{\ast }$ and $b_{\mathrm{sym,surf}}^{\ast }$ are in a
quite good agreement with the phenomenological ones derived from the Weizs%
\"{a}cker mass formula \cite{bomo1}. The dashed and dotted lines in \textrm{%
Fig. 4} show the results obtained from the extended Thomas-Fermi
approximation using Skyrme forces SkM and SLy230b, see Ref. \cite{kosa08}.

\vspace*{2ex}
%
%  FIGURE 3
%
\includegraphics[width=0.85\textwidth]{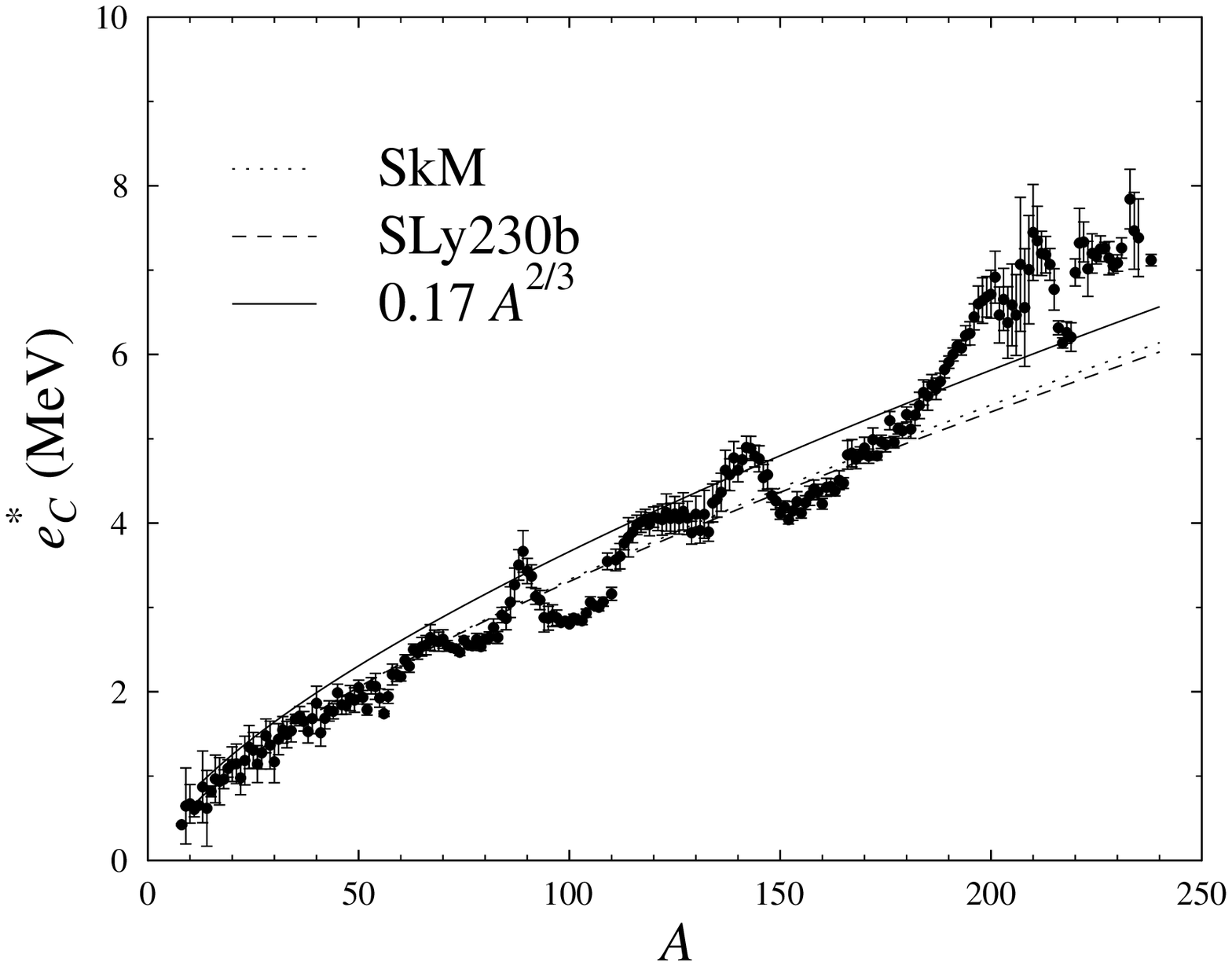}\newline
{\small Fig. 3. The Coulomb energy coefficient $e_{C}^{\ast }(A)$ versus the mass
number $A$ along the beta-stability line. Experimental data are shown as
solid circles. Solid line presets $e_{C}^{\ast }(A)=0.17A^{2/3}$ determined
from the fit to experimental values. Dashed lines are the calculations using
Skyrme forces SkM and SLy230b \cite{kosa08}.}
\vspace*{2ex}

One should pay attention to the sources of the uncertainty in the obtained
values. Firstly, the shell effects in $b_{\mathrm{sym}}^{\ast }$
considerably\ reduce the accuracy of the fit procedure. Secondly, the range
of mass numbers covered by the experimental data does not allow one to
determine $b_{\mathrm{sym,vol}}^{\ast }$ unambiguously, since one cannot
neglect the surface term in (\ref{bi}) even for the large masses ($A\sim 240$%
). Additional uncertainty is because of the small curvature corrections in
the surface symmetry coefficient $b_{\mathrm{sym,surf}}^{\ast }$. Due to the
above, one can obtain quite different symmetry energy coefficients $b_{%
\mathrm{sym,vol}}^{\ast }$ and $b_{\mathrm{sym,surf}}^{\ast }$\ with
approximately the same quality of the fit taking the different intervals of
mass numbers $A$ for the fitting procedure. In particular,\ we have obtained 
$b_{\mathrm{sym,vol}}^{\ast }=32.5$~MeV, $b_{\mathrm{sym,surf}}^{\ast
}=-56.3 $~MeV \ and $r_{S/V}\approx 1.7$\ \ from the fit for $A\geq 50$.%
\textbf{\ }Earlier, a similar feature of the extraction\ of the volume and
the surface terms at the symmetry energy was noted in Ref. \cite{onda04} by
fitting the binding energies of nuclei. Some numerical results are
summarized in the \textrm{Table~1}. Note that in theoretical calculations,
the value of surface-to-volume ratio $r_{S/V}$\ varies strongly within the
interval $1.6\leq r_{S/V}\leq 2.8$, see Refs. \cite%
{list82,moni95,dani03,sawy06}.

\vspace*{1ex}
%
%  FIGURE 4
%
\includegraphics[width=0.85\textwidth]{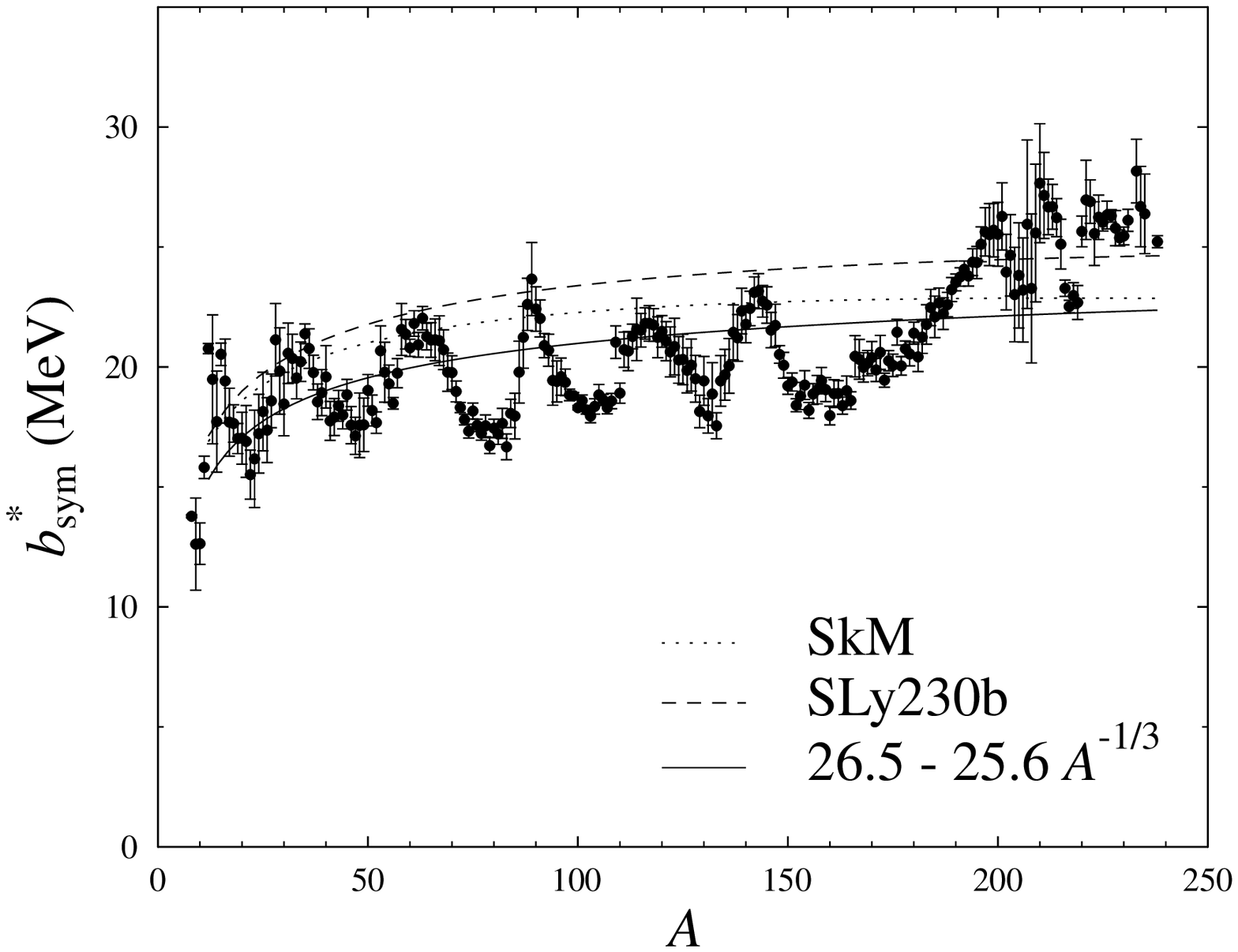}\newline
{\small Fig. 4. The symmetry energy coefficient $b_{\mathrm{sym}}$ as a function of
mass number $A$. Symbols correspond to experimental data. Solid line shows
$b_{\mathrm{sym}}^{\ast}(A)=26.5-25.6A^{-1/3}$ obtained from the
fit to experimental values using Eq.~({\ref{bi}}). Dashed and dotted lines
present the calculations using Skyrme forces SkM and SLy230b, respectively
(see \cite{kosa08}).}
\vspace*{2ex}

\noindent{\small Table 1. Symmetry energy coefficients $b_{\mathrm{sym,vol}}^{\ast }$,
$b_{\mathrm{sym,surf}}^{\ast}$ and the surface-to-volume ratio $r_{S/V}$
obtained from the least square fit of symmetry energy (\ref{bi})
to $b_{\mathrm{sym}}^{\ast}$ from \textrm{Fig. 4} for different fitting
intervals: $A\geq 50$ and $A\geq 12$.}
\begin{center}
\begin{tabular}{cccc}
\hline\hline
& \ \ \ \ $b_{\mathrm{sym,vol}}^{\ast}$, \textrm{MeV}
& \ \ \ \ $b_{\mathrm{sym,surf}}^{\ast }$, \textrm{MeV}
& \ \ \ \ $r_{S/V}$ \\ \hline\hline
$A\geq 50$ & $\ \ \ 32.5$ & $\ \ -56.3$ & $\ \ \ \ 1.73$ \\ 
$A\geq 12$ & $\ \ \ 26.5$ & $\ \ -25.6$ & $\ \ \ \ 1.03$ \\
\hline
\end{tabular}
\end{center}

\textbf{4.} We have established the relation (\ref{lambdaX}) between the
nuclear symmetry energy and the isospin shift of chemical potential $\Delta
\lambda =\lambda _{n}-\lambda _{p}$ that is beyond the beta-stability line.
This relation allowed us to evaluate the symmetry energy independently of
the standard derivation from the mass formula. Moreover this evaluation
performed for different mass number $A$, has been used to derive the
"experimental" values of the $A$-dependent volume, surface and curvature
terms for the isospin symmetry energy.

The presence of shell structure in the single particle levels for neutrons
and protons leads to the characteristic oscillations in $A$-dependencies of
the beta-stability line, the symmetry energy and the Coulomb energy (nuclear
Coulomb radius). We point out the presence of strong amplitude oscillations
of the symmetry energy $b_{\mathrm{sym}}^{\ast }$ which significantly exceed
the corresponding shell effects\ in the binding energy. We do not observe
the existence of exceptionally large values of the symmetry energy
coefficient $b_{\mathrm{sym}}^{\ast }$\ for the mass number $A\approx 100\ $%
which was earlier reported in Ref. \cite{jado03}.

Our approach allowed us to estimate the experimental values of the volume, $%
b_{\mathrm{sym,vol}}^{\ast }$, and the surface, $b_{\mathrm{sym,surf}}^{\ast
}$, contributions to the symmetry energy. The result of the fitting
procedure for $b_{\mathrm{sym,vol}}^{\ast }$\ and $b_{\mathrm{sym,surf}%
}^{\ast }$ depends significantly on the intervals of the mass numbers $A$ \
applied to the fitting procedure.\ By fitting all of the available
information for $A\geq 12$, we have obtained $b_{\mathrm{sym,vol}}^{\ast }=\
26.5$ \textrm{MeV} and the surface contribution $b_{\mathrm{sym,surf}}^{\ast
}=-25.6$ \textrm{MeV} with the surface-to-volume ratio $r_{S/V}\approx 1$\
which is well below the corresponding theoretical results obtained within
the extended Thomas-Fermi approximation (ETFA) with the effective Skyrme
forces in the present work as well as in quantum Skyrme-Hartree-Fock (SHF)
approach \cite{sawy06}. We also point out that the change of the intervals
in mass numbers $A$ for the fitting procedure leads to a significant
difference in the surface contribution $b_{\mathrm{sym,surf}}^{\ast }$ to
the symmetry energy, see rows 2 and 3 in the \textrm{Table 1}. A better
agreement with theoretical results is obtained for the fitting interval with
larger masses $A\geq 50$\ where the leptodermous expansion is more
justified. In this case, the surface-to-volume ratio is given by $%
r_{S/V}\approx 1.7$ which is close to the theoretical result from Ref. \cite%
{sawy06}.

\end{document}